\documentclass[12pt,draftcls,onecolumn]{IEEEtran}
\IEEEoverridecommandlockouts
\usepackage{amsfonts,amsmath,amssymb} % assumes amsmath package installed

\usepackage{psfrag,color}
\usepackage{enumerate,cite,latexsym,graphicx}
\newtheorem{theorem}{Theorem}

\newtheorem{definition}{Definition}

\title{\LARGE \bf Singular Perturbation Approximations for a Class of 
  Linear   Quantum  Systems
}

\author{Ian R.~Petersen%
\thanks{This work was supported by the
Australian Research Council (ARC) and Air Force Office of Scientific
Research (AFOSR). This material is based on research sponsored by the
Air Force Research Laboratory, under agreement number
FA2386-09-1-4089.  The U.S. Government is authorized to reproduce and
distribute reprints for Governmental purposes notwithstanding any
copyright notation thereon.
The views and conclusions contained herein are those of the authors
and should not be interpreted as necessarily representing the official
policies or endorsements, either expressed or implied, of the Air
Force Research Laboratory or the U.S. Government. A Preliminary version of this paper appeared in the Proceedings of the 2010 American Control Conference.}%
\thanks{Ian R. Petersen is with the School of Information Technology and Electrical Engineering, 
        University of New South Wales at the Australian Defence Force Academy, Canberra ACT 2600, Australia.
         {\tt\small i.r.petersen@gmail.com} } }%

\begin{document}

\maketitle
\thispagestyle{empty}
\pagestyle{empty}

\begin{abstract}
This paper considers the use of singular perturbation approximations
for a class of linear  quantum systems arising in the area of linear 
quantum optics. The paper presents results on the physical realizability properties  of the approximate system arising from  singular perturbation model reduction. 
\end{abstract}

%%%%%%%%%%%%%%%%%%%%%%%%%%%%%%%%%%%%%%%%%%%%%%%%%%%%%%%%%%%%%%%%%%%%%%%%%%%%%%%%
\section{Introduction} \label{sec:intro}
The modelling and control of quantum linear systems is an important emerging application area  which is motivated by the fact that quantum mechanical features  emerge as the systems being controlled approach sub-nanometer scales and as the required levels of accuracy in control and estimation  approach quantum noise limits. 
In recent years, there has been considerable interest in the feedback
control and modeling of
linear quantum systems; e.g., see
\cite{DJ99,YK03A,YK03B,BE08,JNP1,NJP1,GGY08,PET08A,NJD09,GJ09,GJ09A,WM10,JG10,NUR10,NUR10A,MaP3,MaP4}.
Such linear quantum  
systems commonly arise in the area of quantum optics; e.g., see
\cite{WM08,GZ00,BR04}. The feedback control of quantum optical systems has
applications in areas such as quantum communications, quantum 
teleportation, and gravity wave detection. In particular, the papers
\cite{PET08A,MaP3,MaP4,NUR10A} have been concerned with a class of linear quantum
systems in which the system can be defined in terms of a set of linear
 quantum stochastic differential equations (QSDEs) expressed
purely in terms of annihilation operators. Such linear  quantum
systems correspond to optical systems made up of passive optical
components  such as optical cavities, beam-splitters, and phase
shifters. The main results of this paper apply to this class of linear quantum systems for the square case in which the number of outputs is equal to the number of inputs. 

This paper is concerned with the use of singular
perturbation approximations in order to obtain reduced dimension models 
for the
class of linear  quantum systems  under consideration.  Singular perturbation approximations are is widely used for obtaining reduced dimension models for classical  systems; e.g., see \cite{KOS76}. In the case of quantum systems, a reduced dimension model may be desired for a quantum plant to be controlled in order to simplify the controller design process which can be very complicated using existing quantum controller design methods such as the quantum LQG method of \cite{NJP1}.  Another application of model reduction for linear quantum systems arises in the case of controller reduction where a reduced dimension controller is obtained from a high order synthesized controller. In the case of coherent quantum control such as considered in \cite{JNP1,NJP1,MaP4}, the controller is required to be a quantum system itself and thus the reduced dimension system must be physically realizable. 

In the physics literature, a commonly used technique in the modeling of quantum systems is the
method of adiabatic elimination, which is closely connected to the
 singular perturbation method in linear systems theory; e.g, see
\cite{GV07,BVS08,BS08,GNW10}. The papers \cite{GV07,BVS08,BS08,GNW10} also consider the issue of convergence of these singular perturbation approximations. In this paper, we consider the properties of the
singular perturbation approximation to a linear quantum system from a linear systems
point of view; e.g., see \cite{KKO86} for a detailed description of
singular perturbation methods in linear systems theory including error characterization in both the time and frequency domains. In particular, we are concerned with the physical realizability properties of the singular perturbation approximation to a linear quantum system. The issue of physical realizability for
linear quantum systems was considered in the papers
\cite{JNP1,NJP1,MaP3,MaP4}. This notion relates to whether a given
QSDE model represents a physical quantum system which obeys the laws
of quantum mechanics. In particular, the results of the papers \cite{JNP1,NJP1,GGY08,MaP3} show that the notion of physical realizability enables a direct connection between results in  quantum linear systems theory and linear systems theory.  In applying singular perturbation methods to
obtain approximate models of quantum systems, it is important that  model obtained is a physically realizable quantum
system so that it retains the essential features of a quantum system. Also, if the approximate model of a quantum plant is to be used for controller synthesis, the controller synthesis procedure may need to exploit the physical realizability of the plant model. In addition, if model order reduction  is applied to a coherent feedback controller which is to be implemented as a quantum system, then this reduced order controller model must be physically realizable.  
% The main result of this paper shows that for the class of
%  linear quantum systems systems under consideration and for the
% considered class of singular perturbations, then the corresponding
% reduced dimension approximate system is guaranteed to be physically
% realizable if the original singularly perturbed system is physically realizable
% for all values of the singular perturbation parameter. 

In the paper \cite{MaP3}, the notion of physical realizability is
shown to be equivalent to the lossless bounded real property for the
class of square linear  quantum systems under consideration. This property requires that the system matrix is Hurwitz and that the system transfer function is unitary for all frequencies. 
The main result of this paper shows that if a singularly perturbed linear quantum system is physically realizable
for all values of the singular perturbation parameter, then the corresponding
reduced dimension approximate system has the property that all of its  poles are in the closed left half of the complex plane and its transfer function is unitary for all frequencies.
These properties indicate that in all but pathological cases, the singular perturbation approximation method will yield a physically realizable reduced dimension system. In addition, an example is given showing one such pathological system in which the singular perturbation approximation is not strictly Hurwitz. 

The paper also presents  a result for a special case of the singularly perturbed linear quantum systems considered in this paper. This special case corresponds to  singular perturbations which arise physically from a perturbation in the system Hamiltonian. In this case, the result shows that  the corresponding
reduced dimension approximate system is always physically realizable. This result can in fact be derived from the nonlinear quantum system results presented in the papers \cite{BVS08,GNW10}. However, we have included this result, along with a straightforward  proof, for the sake of completeness. We have also included an example of a singularly perturbed linear quantum optical system which fits into the subclass of singularly perturbed quantum systems for which this result applies. This example illustrates how such singularly perturbed quantum systems can arise naturally in physical quantum optical systems. 

% Thus, the
% main result of this paper is also related to some results in the
% literature concerning the preservation of the bounded real property
% under singular perturbation model reduction methods; e.g., see
% \cite{MNF97}. However, \cite{MNF97} is concerned with the singular
% perturbation approximation of bounded real systems rather than
% lossless bounded real systems. 

The remainder of this paper proceeds as follows. In Section
\ref{sec:systems}, we define the class of linear  quantum
systems under consideration and recall some preliminary results on
the physical realizability of such systems. In Section
\ref{sec:sing_pert}, we consider the singular perturbation approximation  to a linear
 quantum system. We first  present a result for the class of singularly perturbed linear 
quantum systems under consideration  which relates to the lossless bounded real property. We
then consider a special class of singular perturbations which is
related to corresponding perturbations of the quantum system coupling operator
and Hamiltonian operator. We present a result which relates to
this class of singular perturbations and shows that the corresponding
approximate reduced dimension system is guaranteed to be physically
realizable. In Section \ref{sec:example}, we present a simple example
from the field of quantum optics to illustrate the proposed theory. In
Section \ref{sec:Conclusions}, we present some conclusions.

\section{A Class of Linear  Quantum Systems} \label{sec:systems}
We  consider a class of  linear quantum  systems described in terms of the
 annihilation operator by the following quantum stochastic differential
equations (QSDEs):
\begin{eqnarray}\nonumber
\label{sys}
 da(t) &=& F a(t)dt + G du(t);   \nonumber \\
 dy(t) &=& H a(t)dt + K du(t)
\end{eqnarray}
where $F \in \mathbb{C}^{n \times n}$, $G \in \mathbb{C}^{n
\times m}$, $H \in \mathbb{C}^{m \times n}$ and
$K \in \mathbb{C}^{m \times m}$; e.g., see
\cite{JNP1,GZ00,BR04,MaP3,MaP4}. Here $ a(t) = \left[ {a_1 (t) 
\cdots a_n (t)} \right]^T$ is a vector of (linear combinations of) annihilation operators.
The vector $u(t)$ represents the input signals and is assumed to admit the
decomposition:
\begin{equation}\nonumber
 du(t) = \beta _{u}(t)dt + d\tilde u(t)
\end{equation}
where $\tilde u(t)$ is the noise part of $u(t)$ and $\beta_{u}(t)$
is an adapted process (see \cite{PAR92} and \cite{HP84}).
The noise $\tilde u(t)$ is a vector of quantum noises.  The noise
processes can be represented as operators on an appropriate Fock 
space (for more details see
\cite{PAR92}).
The process $\beta_{u}(t)$ represents variables of other systems
which may be passed to the system (\ref{sys}) via an interaction. More
details concerning this class of quantum systems can be found in the
references \cite{MaP3}, \cite{JNP1}.

\begin{definition} (See \cite{GGY08,MaP3,MaP4}.) 
\label{phys_real}
A  linear quantum 
  system of the form (\ref{sys}) is 
  said to be {\em physically realizable} if there exists a commutation
  matrix $\Theta=\Theta^\dagger > 0$, a coupling matrix $\Lambda$, 
  a Hamiltonian matrix $M=M^\dagger$, and a scattering matrix $S$ such that
\begin{eqnarray}
\label{harmonic}
F  &=& -\Theta \left( { iM + \frac{1}{2}{\Lambda ^\dagger \Lambda}}
\right); \nonumber \\
G  &=& -\Theta \Lambda^\dagger S;\nonumber \\
H  &=& \Lambda;\nonumber \\
K  &=& S
\end{eqnarray}
and $S^\dagger S = I$. 
% In the special case in which the matrix $\Theta
% =I$, the system (\ref{sys}) is said to be canonically physically
% realizable. 
\end{definition} 
Here, the
notation $^\dagger$ represents  conjugate transpose. In this
definition, if the system (\ref{sys}) is physically realizable, then
the matrices $S$, $M$ and $\Lambda$ define a  open harmonic
oscillator with scattering matrix $S$,   coupling operator $ L = \Lambda a $  and
a Hamiltonian operator $\mathcal{H}=a^\dagger M a$; e.g., see 
\cite{GZ00}, \cite{PAR92}, \cite{BHJ07} and
\cite{JNP1}. This definition is an extension of the definition given in
\cite{JNP1,MaP3,MaP4} to allow for a general scattering matrix $S$; e.g.,
see \cite{GGY08}.

% Note that as shown in \cite{MaP3}, if a system of the form (\ref{sys})
% is physically realizable, then there exists a state transformation
% $\tilde a = T a$ such that the resulting transformed system is  canonically physically
% realizable. Also, note that we consider the more general non-canonical case since we allow the elements of $a$ may be linear combinations of annihilation operators rather than simply  annihilation operators.

The following theorem is a straightforward extension of Theorem 5.1 of \cite{MaP3} to
allow for a general scattering matrix $S$. 

\begin{theorem} (See \cite{MaP3}.)
\label{T1}
A  linear quantum system of the form (\ref{sys}) is physically
realizable if and only if there exists a matrix $\Theta=\Theta^\dagger
>0$ such that 
\begin{eqnarray}
\label{physrea}
F\Theta + \Theta F ^\dagger  + GG^\dagger  &=& 0;\nonumber\\
G  &=& -\Theta H^\dagger K;\nonumber\\
K^\dagger K  &=& I.
\end{eqnarray}
In this case, the corresponding Hamiltonian matrix \emph{M} is given by
\begin{equation}\label{Hamiltonian}
M = \frac{i}{2}\left( {\Theta^{-1}F  - F^\dagger\Theta^{-1}
} \right),
\end{equation}
the corresponding coupling matrix $\Lambda$ is given by
\begin{equation}\label{coupling}\Lambda=H \end{equation}
and the corresponding scattering matrix is given by $S=K$. 
\end{theorem}
Note that \emph{M} is a  Hermitian matrix.

\begin{definition}
\label{D2}
The linear  quantum system (\ref{sys}) is said to be \emph{lossless bounded real} if the following conditions hold:
\begin{enumerate}\item[i)] F is a Hurwitz matrix; i.e., all of its eigenvalues have strictly negative real parts; \item[ii)] The transfer function matrix $\Phi(s)=H(sI-F)^{-1}G+K$ satisfies $\Phi(i\omega)^\dagger \Phi(i\omega)=I$ for all $ \omega \in \mathbb{R}.$ \end{enumerate}
\end{definition}

The following definition extends the standard linear systems notion of
minimal realization to linear  quantum systems of the form
(\ref{sys}); see also \cite{MaP3}.
\begin{definition}\label{D3}
A linear  quantum system of the form (\ref{sys}) is said to be \emph{minimal} if the
following conditions hold:
\begin{enumerate}\item[i)] {\em Controllability.} $x^\dagger F=
  \lambda x^\dagger$ for some $\lambda \in \mathbb{C}$ and $x^\dagger
  G =0$ implies $x=0$; \item[ii)] {\em Observability}. $Fx=\lambda
  x$ for some $\lambda \in \mathbb{C}$ and $Hx=0$ implies
  $x=0$.
\end{enumerate}
\end{definition}

The following theorem is an straightforward extension of Theorem 6.6 of \cite{MaP3}  to
allow for a general scattering matrix $S$. 

\begin{theorem}\label{T2} A minimal linear  quantum system of
  the form (\ref{sys}) is physically realizable if and only if  the
  system is lossless bounded real.  
 \end{theorem}
\section{Singularly Perturbed Linear Quantum Systems}
\label{sec:sing_pert}
\subsection{General Singular Perturbations}
We now consider a class of quantum systems of the form (\ref{sys})
dependent on a parameter $\epsilon > 0$ 
which are referred to as singularly perturbed quantum systems:
\begin{eqnarray}
\label{sys1}
da_1(t) &=& F_{11} a_1(t)dt +F_{12} a_2(t)dt + G_1 du(t);   \nonumber\\
da_2(t) &=& \frac{1}{\epsilon}F_{21} a_1(t)dt +\frac{1}{\epsilon}F_{22} a_2(t)dt + \frac{1}{\epsilon}G_2 du(t);   \nonumber\\
 dy(t) &=& H_1 a_1(t)dt +H_2 a_2(t)dt + K du(t).
\end{eqnarray}
This system can be re-written in the more standard singularly
perturbed form (e.g., see \cite{KKO86}):
\begin{eqnarray}\nonumber
\label{sys2}
da_1(t) &=& F_{11} a_1(t)dt +F_{12} a_2(t)dt + G_1 du(t);   \nonumber\\
\epsilon da_2(t) &=& F_{21} a_1(t)dt +F_{22} a_2(t)dt +G_2 du(t);   \nonumber\\
 dy(t) &=& H_1 a_1(t)dt +H_2 a_2(t)dt + K du(t).
\end{eqnarray}
If the matrix $F_{22}$ is non-singular, we can define the
corresponding reduced dimension slow subsystem (e.g., see
\cite{KKO86}) by formally setting $\epsilon =0$ in (\ref{sys2}) to obtain 
\begin{eqnarray}\nonumber
\label{sys3}
 da_1(t) &=& F_0 a_1(t)dt + G_0 du(t);   \nonumber \\
 dy(t) &=& H_0 a_1(t)dt + K_0 du(t)
\end{eqnarray}
where
\begin{eqnarray}
\label{slow_matrices}
F_0 &=& F_{11}-F_{12}F_{22}^{-1}F_{21}; \nonumber \\
G_0 &=& G_{1}-F_{12}F_{22}^{-1}G_{2}; \nonumber \\
H_0 &=& H_{1}-H_{2}F_{22}^{-1}F_{21}; \nonumber \\
K_0 &=& K-H_{2}F_{22}^{-1}G_{2}.
\end{eqnarray}

This is the  singular perturbation approximation to the system (\ref{sys1}). 
We are interested in whether the reduced dimension quantum system
(\ref{sys3}), (\ref{slow_matrices}) is physically realizable if the
singularly perturbed quantum system (\ref{sys2}) is physically
realizable for all $\epsilon > 0$. One approach to addressing this
question might be to apply Theorem \ref{T2} and indeed, we can obtain
the following theorem which is the main result of the paper:

\begin{theorem}\label{T3} If the singularly perturbed linear 
  quantum system  (\ref{sys2}) is physically realizable for all
  $\epsilon > 0$ and the matrix $F_{22}$ is non-singular, then the corresponding reduced dimension quantum
  system (\ref{sys3}), (\ref{slow_matrices}) is such that the matrix
  $F_0$ has all of its eigenvalues in the closed left half of the
  complex plane and the transfer function matrix $\Phi_0(s) =
  H_0(sI-F_0)^{-1}G_0+K_0$ satisfies
\begin{equation}
\label{LBR0}
\Phi_0(i \omega)^\dagger \Phi_0(i \omega) = I
\end{equation}
for all $\omega \in \mathbb{R}$. 
 \end{theorem}
The proof of this theorem is given in the appendix. 

Note that this result is not sufficient to prove the physical
realizability of the reduced dimension quantum system (\ref{sys3}), (\ref{slow_matrices})
since the application of Theorem \ref{T2} requires that the system
realization be minimal and hence the conditions of Theorem \ref{T2} will only be satisfied if the matrix $F_0$ is Hurwitz. However, the properties established in this theorem  indicate that in all but pathological cases, the singular perturbation approximation will yield a physically realizable reduced dimension system. These pathological cases can be detected by testing the eigenvalues and minimality of the reduced order system. In addition, the following  example shows one such pathological system in which the singular perturbation approximation is not strictly Hurwitz and not minimal. 

\noindent
{\bf Example}
We consider a singularly perturbed quantum linear system of the form (\ref{sys2}) where
\begin{eqnarray*}
F_{11} &=& \left[\begin{array}{cc}-\frac{1}{2} & 1\\-1 & -\frac{1}{2}\end{array}\right]; F_{12} = I, F_{21} = \frac{1}{2}I, F_{22} = -I, \nonumber \\
 G_1 &=& -I, G_2 = I, H_1 = I, H_2 = -2 I, K = I. 
\end{eqnarray*}
For each $\epsilon > 0$, we calculate the characteristic polynomial of the matrix 
$F_\epsilon = \left[\begin{array}{cc} F_{11} & F_{12}\\F_{21}/\epsilon & F_{22}/\epsilon 
\end{array}\right]$ to be
\[
p(s) = s^4 + \left(1+\frac{2}{\epsilon}\right)s^3+\left(\frac{5}{4} + \frac{1}{\epsilon^2} + \frac{1}{\epsilon}\right)s^2+\frac{2}{\epsilon}s+\frac{1}{\epsilon^2}. 
\]
From this, it follows using the Routh-Hurwitz criterion that the matrix $F_\epsilon$ is Hurwitz for all $\epsilon > 0$. Furthermore, it is straightforward to verify that the matrix 
$\Theta_\epsilon = \left[\begin{array}{cc}I & 0 \\ 0 & I/\epsilon\end{array}\right] > 0$ satisfies the conditions
\begin{eqnarray}
\label{phys_real_ex}
F_\epsilon \Theta_\epsilon + \Theta_\epsilon F_\epsilon^\dagger + G_\epsilon G_\epsilon^\dagger &=& 0;\nonumber \\
G_\epsilon + \Theta_\epsilon H^\dagger &=& 0 
\end{eqnarray}
where $G_\epsilon = \left[\begin{array}{c}G_1\\G_2/\epsilon\end{array}\right]$ and $H=\left[\begin{array}{cc}H_1 & H_2\end{array}\right]$. Hence, it follows from Theorem \ref{T1} that this singularly perturbed quantum system is physically realizable for all $\epsilon > 0$. Furthermore, it follows from (\ref{phys_real_ex}) that this system is in fact minimal for all $\epsilon > 0$. However, when we consider the  reduced order approximate system, we calculate
$F_0 = \left[\begin{array}{cc}0& 1\\-1 & 0\end{array}\right]$ which is not Hurwitz. Also, $G_0 = 0$, $H_0 =0$, $K_0 = -I$ and thus, the reduced order  system is not minimal. 

% Although the above theorem gives a useful property of singularly
% perturbed quantum systems,  However, it is well known that even if a
% singularly perturbed linear system is controllable and observable for
% all values of the perturbation parameter $\epsilon > 0$, the reduced
% dimension  system need not be controllable or observable;
% e.g., see \cite{CHO77}. 

This example shows that a stronger result than Theorem \ref{T3}, which guarantees minimality and Hurwitzness of the approximate system, cannot be obtained in the general case.
In the next subsection, we consider a special
class of singular perturbations for which the physical realizability
 of the reduced dimension system can be guaranteed.

\subsection{A Special Class of  Singular  Perturbations}

We now consider a special class of singularly perturbed physically
realizable quantum
systems of the form (\ref{sys1}) defined in terms of the matrices $S$, $\Lambda$ and $M$ in
Definition \ref{phys_real}. Indeed, we consider the case in which
$\Theta = I$, 
\[
\Lambda = \left[\begin{array}{cc}\Lambda_1 &
    \frac{1}{\sqrt{\epsilon}}\Lambda_2\end{array}\right];~ 
M = \left[\begin{array}{cc} M_{11} & \frac{1}{\sqrt{\epsilon}}M_{12}\\
\frac{1}{\sqrt{\epsilon}}M_{12}^\dagger & \frac{1}{\epsilon}M_{22}
\end{array}\right]
\]
for all $\epsilon > 0$ where $S^\dagger S = I$, and $M_{11}$ and $M_{22}$ are Hermitian
matrices. Then, substituting these values into (\ref{harmonic}), we
obtain the following linear  quantum system of the form
(\ref{sys}): 
\begin{eqnarray}
\label{sys4}
da_1(t) &=& -\left(\frac{1}{2}\Lambda_1^\dagger\Lambda_1 + i
  M_{11}\right)a_1(t)dt \nonumber \\
&&-\frac{1}{\sqrt{\epsilon}}\left(\frac{1}{2}\Lambda_1^\dagger\Lambda_2 + i
  M_{12}\right)a_2(t)dt -  \Lambda_1^\dagger Sdu(t);   \nonumber\\
da_2(t) &=&-\frac{1}{\sqrt{\epsilon}}\left(\frac{1}{2}\Lambda_2^\dagger\Lambda_1 + i
  M_{12}^\dagger\right)a_1(t)dt \nonumber \\
&&-\frac{1}{\epsilon}\left(\frac{1}{2}\Lambda_2^\dagger\Lambda_2 + i
  M_{22}\right)a_2(t)dt - \frac{1}{\sqrt{\epsilon}} \Lambda_2^\dagger Sdu(t);   \nonumber\\
 dy(t) &=& \Lambda_1 a_1(t)dt +\frac{1}{\sqrt{\epsilon}}\Lambda_2 a_2(t)dt + Sdu(t).
\end{eqnarray}
If we make the  change of variables $\bar a_2(t) =
\frac{1}{\sqrt{\epsilon}}a_2(t)$, this leads to the following singularly
perturbed quantum system of the form (\ref{sys1}):
\begin{eqnarray}
\label{sys5}
da_1(t) &=& -\left(\frac{1}{2}\Lambda_1^\dagger\Lambda_1 + i
  M_{11}\right)a_1(t)dt \nonumber \\
&&-\left(\frac{1}{2}\Lambda_1^\dagger\Lambda_2 + i
  M_{12}\right)\bar a_2(t)dt -  \Lambda_1^\dagger Sdu(t);   \nonumber\\
d\bar a_2(t) &=&-\frac{1}{\epsilon}\left(\frac{1}{2}\Lambda_2^\dagger\Lambda_1 + i
  M_{12}^\dagger\right)a_1(t)dt \nonumber \\
&&-\frac{1}{\epsilon}\left(\frac{1}{2}\Lambda_2^\dagger\Lambda_2 + i
  M_{22}\right)\bar a_2(t)dt - \frac{1}{\epsilon} \Lambda_2^\dagger Sdu(t);   \nonumber\\
 dy(t) &=& \Lambda_1 a_1(t)dt +\Lambda_2 \bar a_2(t)dt + Sdu(t).
\end{eqnarray}
Note that even though we formally let $\epsilon \rightarrow 0$ in the singular perturbation approximation, this state space transformation can be applied for each fixed $\epsilon > 0$.
Then, for the singularly perturbed linear  quantum system (\ref{sys5}), we can
obtain the corresponding reduced dimension approximate system
according to equations (\ref{sys3}), (\ref{slow_matrices}).

The following result is obtained for singularly perturbed linear  quantum systems of the
form (\ref{sys5}). This result can also be derived from the general nonlinear results presented in the papers \cite{BVS08,GNW10}. However, this result for the linear case is included here for the sake of completeness. 

\begin{theorem}
\label{T4}
Consider a singularly perturbed linear  quantum system (\ref{sys5}) which is
physically realizable for all $\epsilon > 0$ and suppose that the
matrix $-\left(\frac{1}{2}\Lambda_2^\dagger\Lambda_2 + i
  M_{22}\right)$ is nonsingular. Then the corresponding
reduced dimension approximate system defined by equations (\ref{sys3}), (\ref{slow_matrices}) is physically
realizable. 
\end{theorem}
The proof of this theorem is given in the appendix. 

\section{Illustrative Example}
\label{sec:example}
\begin{figure}[htbp]
\psfrag{y1}{$y_1$}
\psfrag{u1}{$u_1$}
\psfrag{yt}{$\tilde y$}
\psfrag{at}{$\tilde a$}
\psfrag{gt}{$\tilde \gamma$}
\psfrag{ut}{$\tilde u$}
\psfrag{K1}{$K_1$}
\psfrag{K2}{$K_2$}
\psfrag{y2}{$y_2$}
\psfrag{u2}{$u_2$}
\begin{center}
\includegraphics[width=8cm]{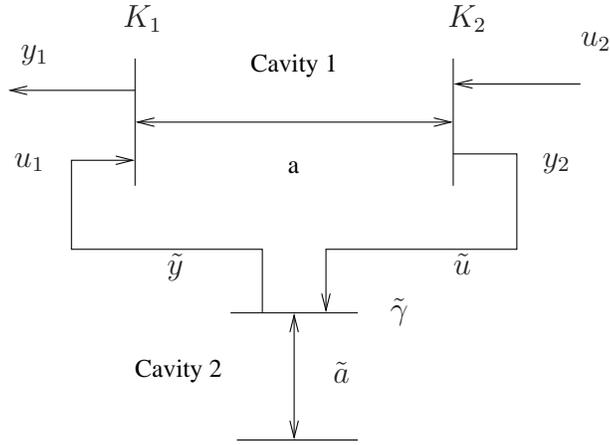}
\end{center}
\caption{A linear optical quantum system.}
\label{F1}
\end{figure}

We consider an example from quantum optics involving the
interconnection of two optical cavities as shown in Figure \ref{F1}. Each
optical cavity consists of two partially reflective mirrors which are
spaced at a specified distance to give a cavity resonant frequency
which corresponds to the frequency of the driving laser; e.g., see \cite{BR04,WM08}.
In practice,
optical isolators would also  need to be included in the optical
connections between the cavities to ensure that the light traveled
only in one direction. 

Here $K_1$ and $K_2$ are the coupling parameters of the first cavity
and $\tilde \gamma$ is the coupling parameter of the second
cavity. These parameters are determined by the physical characteristics
of each cavity including the mirror reflectivities. 
The QSDE of the form (\ref{sys}) describing this quantum
system is as follows:
\begin{eqnarray}
\label{sys7}
d \left[\begin{array}{c}a \\ \tilde a\end{array}\right] &=& 
\left[\begin{array}{cc}-\frac{K_1+K_2}{2} -\sqrt{K_1 K_2} & -\sqrt{K_1
        \tilde \gamma} \\
-\sqrt{K_2  \tilde \gamma} & -\frac{\tilde \gamma}{2}\end{array}\right]
\left[\begin{array}{c}a \\ \tilde a\end{array}\right]dt %\nonumber \\&&
-\left[\begin{array}{c}\sqrt{K_1}+\sqrt{K_2} \\ \sqrt{\tilde
    \gamma} \end{array}\right] du_2; \nonumber \\
dy_1 &=& \left[\begin{array}{cc}\sqrt{K_1}+\sqrt{K_2} & \sqrt{ \tilde
      \gamma}
\end{array}\right]\left[\begin{array}{c}a \\ \tilde
  a\end{array}\right]dt + du_2.
\end{eqnarray}
We wish to consider the reduced dimension approximation to this system
which is obtained by letting $\tilde \gamma \rightarrow
\infty$. This corresponds to the case in which the mirrors in Cavity
2 are perfectly reflecting and so there is a direct optical feedback
from the output $y_2$ of Cavity 1 into the input $u_1$ of Cavity 1. If we
let $\epsilon = \frac{1}{\tilde \gamma}$, it is straightforward to
verify that this system is a system of the form (\ref{sys4}) with
$\Theta = I$, $S=I$, 
$\Lambda_1 = \sqrt{K_1}+\sqrt{K_2}$, $\Lambda_2 = 1$, $M_{11} =0$, $M_{22} =0$,
$M_{12} = \frac{i}{2}\left(\sqrt{K_1}-\sqrt{K_2}\right)$. 
% The fact
% that $\Theta = I$ in this example is to be expected since the system
% variables $a$ and $\tilde a$ satisfy the canonical commutation
% relations; e.g., see \cite{MaP3}. 
With the change of
variables $\bar a = \sqrt{\tilde \gamma} \tilde a =
\frac{1}{\sqrt{\epsilon}} \tilde a$, the system becomes
\begin{eqnarray}
\label{sys8}
d \left[\begin{array}{c}a \\ \ \epsilon \bar a\end{array}\right] &=& 
\left[\begin{array}{cc}-\frac{K_1+K_2}{2} -\sqrt{K_1 K_2} & -\sqrt{K_1} \\
-\sqrt{K_2} & -\frac{1}{2}\end{array}\right]
\left[\begin{array}{c}a \\ \tilde a\end{array}\right]dt% \nonumber \\&&
-\left[\begin{array}{c}\sqrt{K_1}+\sqrt{K_2} \\ 1 \end{array}\right] du_2; \nonumber \\
dy_1 &=& \left[\begin{array}{cc}\sqrt{K_1}+\sqrt{K_2} &1
\end{array}\right]\left[\begin{array}{c}a \\ \tilde
  a\end{array}\right]dt + du_2
\end{eqnarray}
which is a singularly perturbed quantum system of the form
(\ref{sys2}). Hence, the corresponding reduced dimension slow subsystem
(\ref{sys3}), (\ref{slow_matrices}) is given by
\begin{eqnarray}
\label{sys9}
da &=& \left(-\frac{K_1+K_2}{2} + \sqrt{K_1K_2} \right) adt 
%\nonumber\\&&
+
\left(\sqrt{K_1}-\sqrt{K_2}\right) du_2\nonumber \\
dy_1&=&\left(\sqrt{K_1}-\sqrt{K_2}\right)adt - du_2.
\end{eqnarray}
Since the system (\ref{sys8}) satisfies the conditions of Theorem
\ref{T4}, it follows from this theorem that the system (\ref{sys9})
will be physically realizable. This can also be verified directly by
noting that the system (\ref{sys9}) satisfies the conditions of
Theorem \ref{T1} with $\Theta =1$. 

Note that for this example, if $K_1 = K_2$, then the reduced dimension
quantum system is uncontrollable, unobservable and has a pole at the
origin. 
% This indicates that an approach based on Theorem 2 and Theorem
% 3 would not be successful in proving that this system is physically
% realizable. 
\section{Conclusions}
\label{sec:Conclusions}
In this paper, we have considered the physical realizability properties of the 
singular perturbation approximation to a class of singularly perturbed linear quantum systems. These results may be useful in the
modeling of  linear quantum systems such as gravity wave
detectors where a simplified model is required without
sacrificing physical realizability. 
% Future research might be directed
% towards determining whether this result can be extended to more
% general linear quantum systems and more general classes of singular
% perturbations. 

\section*{acknowledgment}
The author is wishes to acknowledge useful research discussions with Matthew
James, Elanor
Huntington, Michele Heurs and Hendra Nurdin.
\section*{Appendix}
\subsection*{Proof of Theorem \ref{T3}.}
If the singularly perturbed quantum system (\ref{sys2}) is physically
realizable for all $\epsilon > 0$, then it follows from Theorem
\ref{T1} that for all $\epsilon > 0$, there exists a matrix $\Theta >
0$ such that the matrices
\begin{eqnarray*}
F_\epsilon &=& \left[\begin{array}{cc} F_{11} & F_{12} \\
    \frac{1}{\epsilon}F_{21} &
    \frac{1}{\epsilon}F_{22}\end{array}\right];~
G_\epsilon = \left[\begin{array}{c} G_{1}\\
    \frac{1}{\epsilon}G_{2}\end{array}\right];\nonumber \\
H &=& \left[\begin{array}{cc} H_{1} & H_{2} \end{array}\right];~K
\end{eqnarray*}
satisfy the conditions (\ref{physrea}). Hence, it follows from the
first of these equalities and Fact 12.21.3 of \cite{BER05} that matrix
$F_\epsilon$ has all of its eigenvalues in the closed left half of the
complex plane for all $\epsilon > 0$. Then, using a standard result on
singularly perturbed linear systems (e.g., see Theorem 3.1 on page 57 of
\cite{KKO86}) it follows that the matrix $F_0$ has all of its eigenvalues in the closed left half of the
complex plane.  

With the matrices $F_\epsilon$, $G_\epsilon$, $H$ and $K$ defined as
above,  it follows by a straightforward but tedious calculation
that we can write the transfer function $\Phi_\epsilon(s) =
H(sI-F_\epsilon)^{-1}G_\epsilon + K$ in the form:
\begin{eqnarray*}
\Phi_\epsilon(s) &=& \left(H_0 + H_2
  F_{22}^{-1}\left(I-\frac{F_{22}}{\epsilon
      s}\right)^{-1}F_{21}\right)
%\nonumber \\&& \times 
\left[sI-F_0-F_{12}\left(I-\frac{F_{22}}{\epsilon
        s}\right)^{-1}F_{21}\right]^{-1}\nonumber \\
&& \times \left(G_0+F_{12}\left(I-\frac{F_{22}}{\epsilon
        s}\right)^{-1}G_2\right)\nonumber \\
&& +K_0+H_2F_{22}^{-1}\left(I-\frac{F_{22}}{\epsilon
      s}\right)^{-1}G_2
\end{eqnarray*}
where the matrices $F_0$, $G_0$, $H_0$, $K_0$ are defined as in
(\ref{slow_matrices}). 

Now for small values of $\epsilon > 0$, we can approximate the term $\left(I-\frac{F_{22}}{\epsilon
      s}\right)^{-1}$ in the above expression as follows:
\[
\left(I-\frac{F_{22}}{\epsilon
      s}\right)^{-1} =-\epsilon s F_{22}^{-1} + O(\epsilon^2).
\]
From this, it follows that we can write
\begin{eqnarray*}
\Phi_\epsilon(s) &=& \left(H_0 -\epsilon s H_2
  F_{22}^{-2}F_{21}\right)\left(sI-F_0\right)^{-1}
%\nonumber \\&& \times 
\left[I+\epsilon s
  F_{12}F_{22}^{-1}F_{21}\left(sI-F_0\right)^{-1}\right]^{-1}
\nonumber \\ 
&& \times \left(G_0-\epsilon s
  F_{12}F_{22}^{-1}G_2\right)\nonumber \\
&& +K_0- \epsilon sH_2F_{22}^{-2}G_2 + O(\epsilon^2).
\end{eqnarray*}
From this and some further straightforward manipulations and
simplifications, we can obtain
\begin{eqnarray}
\label{TF_pert}
\Phi_\epsilon(s) &=& \Phi_0(s) 
%\nonumber \\&&
- \epsilon s
\left(H_0\left(sI-F_0\right)^{-1}F_{12}+H_2F_{22}^{-1}\right)F_{22}^{-1}
%\nonumber \\&& \times 
\left(F_{21}\left(sI-F_0\right)^{-1}G_0+G_2\right)\nonumber \\
&&+ O(\epsilon^2).
\end{eqnarray}

Now using the fact that the matrices $F_\epsilon$, $G_\epsilon$, $H$ and $K$ satisfy
the conditions (\ref{physrea}), we will show that transfer function matrix $\Phi_\epsilon(s)$is unitary at all frequencies.  Indeed using (\ref{physrea}), we have for all $\omega \in \mathbb{R}$
\begin{eqnarray*}
\lefteqn{H\left(i\omega I - F_\epsilon\right)^{-1}G_\epsilon 
G_\epsilon^\dagger\left(-i\omega I - F_\epsilon^\dagger\right)^{-1}H^\dagger} \nonumber \\
&& =H\left(i\omega I - F_\epsilon\right)^{-1}
\left[\left(i\omega I-F_\epsilon\right)\Theta
+\Theta \left(-i\omega I -F_\epsilon^\dagger\right) \right]\left(-i\omega I - F_\epsilon^\dagger\right)^{-1}H^\dagger\nonumber \\
&&=H\Theta\left(-i\omega I - F_\epsilon^\dagger\right)^{-1}H^\dagger
+H\left(i\omega I - F_\epsilon\right)^{-1}\Theta H^\dagger \nonumber \\
&&=-KG_\epsilon^\dagger\left(-i\omega I - F_\epsilon^\dagger\right)^{-1}H^\dagger
- H\left(i\omega I - F_\epsilon\right)^{-1}G_\epsilon K^\dagger. 
\end{eqnarray*}
Now using the third equation of (\ref{physrea}) and the fact that $K$ is square, we have for all $\omega \in \mathbb{R}$
\begin{eqnarray*}
0&=& I - K K^\dagger -KG_\epsilon^\dagger\left(-i\omega I - F_\epsilon^\dagger\right)^{-1}H^\dagger
- H\left(i\omega I - F_\epsilon\right)^{-1}G_\epsilon K^\dagger \nonumber \\
&&- H\left(i\omega I - F_\epsilon\right)^{-1}G_\epsilon 
G_\epsilon^\dagger\left(-i\omega I - F_\epsilon^\dagger\right)^{-1}H^\dagger \nonumber \\
&=& I - \Phi_\epsilon(i \omega)\Phi_\epsilon(i \omega)^\dagger.
\end{eqnarray*}
Therefore, since $\Phi_\epsilon(i \omega)$ is square we have
\begin{equation}
\label{unitary_e}
\Phi_\epsilon(i \omega)^\dagger \Phi_\epsilon(i \omega) = I~\forall
\omega \in \mathbb{R}
\end{equation}
for all $\epsilon > 0$. Hence, it follows from (\ref{TF_pert}) and the fact that (\ref{unitary_e}) holds for all $\epsilon > 0$ that we must have
\[
\Phi_0(i \omega)^\dagger \Phi_0(i \omega) = I
\]
for all $\omega \in \mathbb{R}$. 
This completes the proof of the theorem. $\Box$

\subsection*{Proof of Theorem \ref{T4}.}
For the singularly perturbed linear  quantum system (\ref{sys5}), it is
straightforward but tedious to verify that the corresponding
reduced dimension slow subsystem  (\ref{sys3}), (\ref{slow_matrices}) is
given by 
\begin{eqnarray}
\label{sys6}
 da_1(t) &=& -\left(i \tilde M+\frac{1}{2}\tilde \Lambda^\dagger\tilde\Lambda\right)a_1(t)dt - \tilde \Lambda^\dagger \tilde S du(t);   \nonumber \\
 dy(t) &=& \tilde\Lambda a_1(t)dt + \tilde S du(t)
\end{eqnarray}
where
\begin{eqnarray}
\label{matrices}
\tilde\Lambda &=& \Lambda_1 %\nonumber \\&&
- \Lambda_2\left(\frac{1}{2}\Lambda_2^\dagger\Lambda_2 + i
  M_{22}\right)^{-1}\left(\frac{1}{2}\Lambda_2^\dagger\Lambda_1 + i
  M_{12}^\dagger\right);\nonumber \\
\tilde S &=& S - \Lambda_2\left(\frac{1}{2}\Lambda_2^\dagger\Lambda_2 + i
  M_{22}\right)^{-1}\Lambda_2^\dagger S;\nonumber \\
\tilde M &=& M_{11} %\nonumber \\&&
+ \frac{1}{4} \Lambda_1^\dagger \Lambda_2\left(\frac{1}{2}\Lambda_2^\dagger\Lambda_2 - i
  M_{22}\right)^{-1}M_{22}
%\nonumber \\&& \times 
\left(\frac{1}{2}\Lambda_2^\dagger\Lambda_2 + i
  M_{22}\right)^{-1}\Lambda_2^\dagger\Lambda_1\nonumber \\
&&+ M_{12}\left(\frac{1}{2}\Lambda_2^\dagger\Lambda_2 - i
  M_{22}\right)^{-1}M_{22}%\nonumber \\&& \times 
\left(\frac{1}{2}\Lambda_2^\dagger\Lambda_2 + i
  M_{22}\right)^{-1}M_{22}^\dagger\nonumber \\
&&-\frac{1}{4} M_{12}\left(\frac{1}{2}\Lambda_2^\dagger\Lambda_2 - i
  M_{22}\right)^{-1}\Lambda_2^\dagger\Lambda_2%\nonumber \\&& \times 
\left(\frac{1}{2}\Lambda_2^\dagger\Lambda_2 + i
  M_{22}\right)^{-1}\Lambda_2^\dagger\Lambda_1\nonumber \\
&&-\frac{1}{4} \Lambda_1^\dagger \Lambda_2\left(\frac{1}{2}\Lambda_2^\dagger\Lambda_2 - i
  M_{22}\right)^{-1}\Lambda_2^\dagger\Lambda_2
%\nonumber \\&& \times 
\left(\frac{1}{2}\Lambda_2^\dagger\Lambda_2 + i
  M_{22}\right)^{-1}M_{12}^\dagger.
\end{eqnarray}
Furthermore, it is straightforward to verify that $\tilde S^\dagger
\tilde S = I$
and the matrix $\tilde M$ is Hermitian. Hence, it follows from 
Definition \ref{phys_real} that the system (\ref{sys6}) is physically realizable with the matrices 
$\tilde \Lambda$, $\tilde S$, $\tilde M$ defined as above and with $\Theta = I$. This
completes the proof of the theorem.
$\Box$

% \bibliography{/home/irp/Bibliog/irpnew}  
% \bibliographystyle{IEEEtran}

\end{document}